\documentclass[submission,copyright,creativecommons]{eptcs}
\providecommand{\event}{NCMA 2022} 
\usepackage[strings]{underscore}           

\usepackage{amssymb}
\usepackage{amsmath}
\usepackage{amsfonts}
\usepackage[all]{xy}

\newcommand{\Accept}{{\sf Accept}}
\newcommand{\Reject}{{\sf Reject}}

\newcommand{\N}{\mathbb{N}}

\newcommand{\beginproof}{{\noindent \bf Proof.~}}
\newcommand{\myendproof}{\hspace*{\fill} $\Box$ \vspace{+0.2cm}}

\newtheorem{lemma}{Lemma}{\bf}{\it}
\newtheorem{proposition}[lemma]{Proposition}{\bf}{\it}
\newtheorem{theorem}[lemma]{Theorem}{\bf}{\it}
\newtheorem{corollary}[lemma]{Corollary}{\bf}{\it}
\newtheorem{definition}[lemma]{Definition}{\bf}{\it}
\newtheorem{example}[lemma]{Example}{\bf}{\rm}
{\bf}{\rm}
{\bf}{\it}

\title{Non-Returning Finite Automata With Translucent Letters}
\author{Franti{\v s}ek Mr\'az
\institute{Charles University\\
Department of Computer Science\\
Malostransk\'e n\'am.~25\\
118 00 PRAHA, Czech Republic}
\email{frantisek.mraz@mff.cuni.cz}
\and
Friedrich Otto
\institute{Universit\"at Kassel\\
Fachbereich Elektrotechnik/Informatik\\
34109 KASSEL, Germany}
\email{\quad f.otto@uni-kassel.de}
}

\def\titlerunning{Non-Returning NFAs with Translucent Letters}
\def\authorrunning{F.~Mr\'az and F.~Otto}

\begin{document}

\maketitle

\begin{abstract}
Here we propose a variant of the nondeterministic finite automaton with translucent letters (NFAwtl)
which, after reading and deleting a letter, does not return to the left end of its tape,
but rather continues from the position of the letter just deleted.
When the end-of-tape marker is reached, our automaton can decide whether to accept, to reject, or to continue,
which means that it again reads the remaining tape contents from the beginning.
This type of automaton, called a \emph{non-returning finite automaton with translucent letters} or an \emph{nrNFAwtl},
is strictly more expressive than the NFAwtl.
We study the expressive capacity of this type of automaton and that of its deterministic variant.
Also we are interested in closure properties of the resulting classes of languages and in decision problems.
\end{abstract}

\section{Introduction}\label{sec0}
While a (deterministic or nondeterministic) finite automaton reads its input strictly from left to right, letter by letter,
by now many types of automata have been considered in the literature that process their inputs in a different, more involved way.
Under this aspect, the most extreme is the \emph{jumping finite automaton} of Meduna and Zemek~\cite{MeZe2012} (see also~\cite{FPS2012}),
which, after reading a letter, jumps to an arbitrary position of the remaining input.
It is known that the jumping finite automaton accepts languages that are not even context-free,
like the language $\{\,w\in\{a,b,c\}^*\mid |w|_a = |w|_b = |w|_c\,\}$,
but at the same time, it does not even accept the finite language~$\{ab\}$.

Another example is the \emph{nondeterministic linear automaton} (or {NLA}) studied by Loukanova in~\cite{Lou2007},
which is a nondeterministic finite automaton with two heads,
one reading the input from left to right,
the other reading the input from right to left.
This model can be simulated by a model with one head that reads alternatingly the first and the
last letter.
It is easily seen that this model characterizes the class {\sf LIN} of linear context-free languages.
Actually, the NLA corresponds to the \emph{$5'\to 3'$-sensing Watson-Crick automaton} defined by Nagy in~\cite{Nag2007}.

Moreover, there is the \emph{restarting automaton} as introduced by Jan{{\v c}}ar, Mr{\'a}z, Pl{\'a}tek, and Vogel in~\cite{JMPV95},
which processes a given input in cycles, in each cycle scanning the remaining input from left to right until
it deletes one or more letters, returns its head to the left end of the remaining input,
and reenters its initial state.
If using a head of size larger than one, these so-called R-automata accept a proper superclass of the regular languages
that is incomparable to the context-free and the growing context-sensitive languages (see, e.g.,~\cite{otto139}),
while with a head of size one, they accept exactly the regular languages~\cite{Mra01}.

Finally, there is the (deterministic and nondeterministic) finite automaton \emph{with translucent letters} (or DFAwtl and NFAwtl) of Nagy and Otto~\cite{otto185},
which is equivalent to a cooperating distributed system of stateless deterministic R-automata with heads of size one.
For each state $q$ of an NFAwtl, there is a set $\tau(q)$ of \emph{translucent letters},
which is a subset of the input alphabet that contains those letters that the automaton cannot see when it is in state~$q$.
Accordingly, in each step, the NFAwtl just reads (and deletes) the first letter from the left that it can see,
that is, that is not translucent for the current state.
It has been proved that the NFAwtl accepts a class of semi-linear languages
that properly contains all rational trace languages,
while its deterministic variant, the DFAwtl,
is properly less expressive.
In fact, the DFAwtl just accepts a class of languages that is incomparable to the rational trace languages with respect to inclusion~\cite{otto176,NaOtLATA2011,otto195,otto206}.
In addition, while the obvious upper bound for the time complexity of the membership problem for a DFAwtl is ${\rm DTIME}(n^2)$,
a better upper bound of ${\rm DTIME}(n \cdot \log n)$ is derived in~\cite{NagKov14}.

Here we propose a variant of the nondeterministic finite automaton with translucent letters
which, after reading and deleting a letter, does not return to the left end of its tape,
but that rather continues from the position of the letter just deleted.
When the end-of-tape marker is reached, our automaton can decide whether to accept, reject or continue,
which means that it again reads the remaining tape contents from the beginning.
We prove that this type of automaton, called a \emph{non-returning finite automaton with translucent letters} or an \emph{nrNFAwtl},
is strictly more expressive than the NFAwtl.
However, as we shall see,
its deterministic variant, the \emph{nrDFAwtl}, which is more expressive than the DFAwtl,
is still not powerful enough to accept all rational trace languages.
In this paper, we concentrate on the problem of determining just how expressive these types of automata are
and on the complexity of their membership problems,
but we are also interested in closure and non-closure properties of the resulting classes of languages.

This paper is structured as follows.
In Section~\ref{sec1}, we present the formal definition of the non-returning finite automaton with translucent letters,
we explain its workings by a detailed example, and we derive a kind of normalized form for this type of automaton.
In the next section, which is the main part of the paper,
we compare the classes of languages that are accepted by the nondeterministic and the deterministic non-returning finite automaton with translucent letters
to the language classes accepted by the DFAwtl and the NFAwtl, to the rational trace languages, and to the classes of the Chomsky hierarchy,
establishing some proper inclusion results and some incomparability results.
Then, in Section~\ref{sec3},
we present a few closure and non-closure properties for the classes of languages that are accepted by the nondeterministic and the deterministic
non-returning finite automaton with translucent letters.
Finally, in Section~\ref{sec4}, we study the complexity of the membership problem for the nrDFAwtl,
showing that it is decidable in time ${\mathrm O}(n\cdot (\log n)^2)$ whether a word of length $n$ is accepted by a given nrDFAwtl.
In the concluding section, we summarize our results
and state a number of open problems for future work.

\section{Definitions}\label{sec1}
In order to use it as a reference, we restate the definition of the nondeterministic finite automaton with translucent letters from~\cite{otto185}.

\begin{definition}\label{DefNFAwtl1}
A \emph{finite automaton with translucent letters},
an \emph{NFAwtl}\index{NFAwtl} for short,
is defined as a 7-tuple $A = (Q,\Sigma,\lhd,\tau,I,F,\delta)$,
where $Q$ is a finite set of internal states,
$\Sigma$ is a finite alphabet of input letters,
$\lhd\not\in\Sigma$ is a special symbol that is used as an \emph{end-of-tape marker},
$\tau:Q\to \mathcal{P}(\Sigma)$ is a \emph{translucency mapping},
$I\subseteq Q$ is a set of initial states,
$F\subseteq Q$ is a set of final states,
and $\delta:Q\times\Sigma\to \mathcal{P}(Q)$ is a \emph{transition relation}.
Here it is required that, for each state $q\in Q$ and each letter $a\in \Sigma$,
if $a\in\tau(q)$,
then $\delta(q,a)=\emptyset.$
For each state $q\in Q$, the letters from the set $\tau(q)$ are \emph{translucent} for~$q$,
that is, in state $q$ the automaton $A$ does not see these letters.
\end{definition}

An {NFAwtl} $A = (Q,\Sigma,\lhd,\tau,I,F,\delta)$ works as follows.
For an input word $w\in\Sigma^*$,
it starts in a nondeterministically chosen initial state
$q_0\in I$ with the word $w\cdot\lhd$ on its tape.
Assume that $w= a_1a_2\cdots a_n$ for some $n\ge 1$ and $a_1,a_2,\ldots,a_n\in\Sigma$,
and assume that $A$ is in state~$q\in Q$.
Then $A$ looks for the first occurrence from the left of a letter
that is not translucent for state~$q$,
that is, if $w=uav$ such that $u\in(\tau(q))^*$ and $a\not\in \tau(q)$,
then $A$ nondeterministically chooses a state $q_1\in\delta(q,a)$,
erases the letter $a$ from the tape, thus producing the tape contents $uv\cdot\lhd$,
its internal state is set to~$q_1$, the head returns to the first letter on the tape,
and the computation continues.
In case $\delta(q,a)=\emptyset$,
$A$ halts without accepting.
Finally, if $w\in(\tau(q))^*$,
then $A$ reaches the end-of-tape marker~$\lhd$ and the computation halts.
In this case, $A$ accepts if $q$ is a final state;
otherwise, it does not accept.
Thus, $A$ executes the following computation relation on its set
$Q\cdot\Sigma^*\cdot\lhd\,\cup\,\{\Accept,\Reject\}$ of configurations:
$$qw\cdot\lhd \vdash_A \left\{ \begin{array}{ll}
     q'uv\cdot\lhd, & \mbox{if }w=uav,\,u\in(\tau(q))^*,\,a\not\in\tau(q),\mbox{ and }q'\in\delta(q,a),\\
      \Reject,       & \mbox{if }w=uav,\, u\in(\tau(q))^*,\,
      a\not\in\tau(q_0),\mbox{ and }\delta(q,a)=\emptyset,\\
     \Accept,      & \mbox{if }w\in(\tau(q))^* \mbox{ and }q\in F,\\
     \Reject,      & \mbox{if }w\in(\tau(q))^*\mbox{ and }q\not\in F.
     \end{array}\right.
$$
A word $w\in\Sigma^*$ is \emph{accepted by} $A$ if there exists an initial state $q_0\in I$
and a computation $q_0w\cdot \lhd \vdash_A^* \Accept$, where
$\vdash_A^*$ denotes the reflexive transitive closure of the above single-step
computation relation~$\vdash_A$.
Now $L(A) = \{\,w\in\Sigma^*\mid w\mbox{ is accepted by }A\,\}$
is the \emph{language accepted by}~$A$ and
$\mathcal{L}({\sf NFAwtl})$ denotes the class of all languages
that are accepted by~NFAwtls.

\begin{definition}\label{DefDFAwtl}
An NFAwtl $A= (Q,\Sigma,\lhd,\tau,I,F,\delta)$ is a
\emph{deterministic finite automaton with translucent letters},
abbreviated as \emph{DFAwtl},
if $|I|=1$ and
if $|\delta(q,a)|\le 1$ for all $q\in Q$ and all $a\in\Sigma$.
Then $\mathcal{L}({\sf DFAwtl})$ denotes the class of all languages
that are accepted by DFAwtls.
\end{definition}

For future reference, we present an example of a DFAwtl.

\begin{example}\label{ExDFAwtl}
Let $A_c = (Q,\Sigma,\lhd,\tau,q_0,F,\delta)$ be the DFAwtl
that is given through $Q=\{q_0,q_a,q_{a}',q_b,q_b'\}$,
$\Sigma = \{a,b,a',b'\}$, $F=\{q_0\}$ and the functions $\tau$ and $\delta$ that are defined as follows:
$$\begin{array}{lcllcllcllcl}
\tau(q_0) &  = & \emptyset, & \tau(q_a) & = & \{a',b'\}, & \tau(q_a') & = & \{a,b\},\\
          &    &            & \tau(q_b) & = & \{a',b'\}, & \tau(q_b') & = & \{a,b\},\\[+1mm]
\delta(q_0,a) & = & q_a', & \delta(q_0,b) & = & q_b',&
\delta(q_0,a') & = & q_a, & \delta(q_0,b') & = & q_b,\\
\delta(q_a,a)  & = & q_0, & \delta(q_b,b) & = & q_0, &
\delta(q_a',a')  & = & q_0, & \delta(q_b',b') & = & q_0,\\
\end{array}$$
and $\delta(q_a,b), \delta(q_b,a), \delta(q_a',b')$, and $\delta(q_b',a')$ are undefined.
For the word $abba'b'ab'a'$, $A_c$ executes the following accepting computation:
$$\begin{array}{ccccccccc}
q_0abba'b'ab'a'\lhd & \vdash_{A_c} & q_a'bba'b'ab'a'\lhd & \vdash_{A_c} & q_0bbb'ab'a'\lhd & \vdash_{A_c} & q_b'bb'ab'a'\lhd \\
                    & \vdash_{A_c} & q_0bab'a'\lhd       & \vdash_{A_c} & q_b'ab'a'\lhd      & \vdash_{A_c} & q_0aa'\lhd\\
                    & \vdash_{A_c} & q_a'a'\lhd          & \vdash_{A_c} & q_0\lhd            & \vdash_{A_c} & \Accept.
  \end{array}$$
In fact, if $\varphi$ denotes the morphism that is defined through $\varphi(a) = a'$ and $\varphi(b) = b'$,
then it is easily checked that $L(A_c) = \{\,{\mathrm{sh}}(w,\varphi(w)) \mid w\in\{a,b\}^*\,\}$,
where ${\mathrm{sh}}$ denotes the shuffle operation.
\hspace*{\fill}$\blacksquare$
\end{example}

The above language $L(A_c)$ is not context-free. In fact, it is not even a growing context-sensitive language.
Let $\pi:\{a,b,a',b'\}^* \to \{a,b\}^*$ be the morphism that is defined through
$a\mapsto a$, $b\mapsto b$, $a'\mapsto a$, and $b'\mapsto b$.
Then
$$\begin{array}{lclcl}
\pi(L(A_c)\,\cap\,(\{a,b\}^*\cdot \{a',b'\}^*)) & = \pi(\{\,w\varphi(w)\mid w\in\{a,b\}^*\,\})
                                                & = & \{\,ww\mid w\in \{a,b\}^*\,\},
\end{array}$$
which is the copy language on $\{a,b\}^*$ that is not growing context-sensitive~\cite{otto84}.
As the class {\sf GCSL} of growing context-sensitive languages is closed under the operations of intersection
with regular sets and non-erasing morphisms, this implies that the language $L(A_c)$ is not growing context-sensitive, either.
Thus, Example~\ref{ExDFAwtl} shows that already DFAwtls accept quite some complicated languages in comparison to the Chomsky hierarchy.
On the other hand, the language $L(A_c)$ is the rational trace language that is obtained
from the regular language $\{aa',bb'\}^*$ through the dependency relation
$D=\{(a,b),(b,a),(a',b'),(b',a')\}$ (see, e.g.,~\cite{otto176,otto195}).
\vspace{+2mm}

As defined above, an NFAwtl performs each step of its computation starting from the first letter on its tape:
it looks for the first letter that is not translucent for the current state,
deletes it, changes its state, and returns to the first letter.
Here we propose a variant of this type of automaton that does not necessarily return to the first letter,
but that continues from the position of the letter deleted, returning to the first letter only after
the tape contents has been scanned completely.
Next, we present the formal definition of this type of automaton,
which is called the \emph{non-returning finite automaton with translucent letters}
or \emph{nrNFAwtl} for short.

\begin{definition}\label{DefnrNFAwtl}
An \emph{nrNFAwtl} is defined by a 6-tuple $A = (Q,\Sigma,\lhd,\tau,I,\delta)$,
where $Q$ is a finite set of internal states,
$\Sigma$ is a finite alphabet of input letters,
$\lhd\not\in\Sigma$ is a special symbol that is used as an \emph{end-of-tape marker},
$\tau:Q\to \mathcal{P}(\Sigma)$ is a \emph{translucency mapping},
$I\subseteq Q$ is a set of initial states,
and $$\delta:Q\times(\Sigma\cup\{\lhd\})\to (\mathcal{P}(Q)\cup\{\Accept\})$$ is a \emph{transition relation}.
Here it is required that, for each state $q\in Q$ and each letter $a\in \Sigma$,
$\delta(q,a)\subseteq Q$, and
if $a\in\tau(q)$,
then $\delta(q,a)=\emptyset$.
For each state $q\in Q$, the letters from the set $\tau(q)$ are \emph{translucent} for~$q$,
that is, in state $q$ the automaton $A$ does not see these letters.
\end{definition}

From the above definition, we see that $\delta(q,\lhd)$ is either a subset of $Q$ or the operation $\Accept$,
that is, on seeing the end-of-tape marker $\lhd$ in state~$q$,
the nrNFAwtl $A$ has either the option to change its state or to accept.
The {nrNFAwtl} $A = (Q,\Sigma,\lhd,\tau,I,\delta)$ works as follows.
For an input word $w\in\Sigma^*$,
$A$ starts in a nondeterministically chosen initial state
$q_0\in I$ with the word $w\cdot\lhd$ on its tape.
This situation is described by the configuration $q_0w\cdot\lhd$.
Now assume that $A$ is in a configuration of the form $xq_1w\cdot\lhd$,
where $q_1\in Q$ and $x,w\in \Sigma^*$,
that is, $A$ is in state $q_1$, the tape contains the word $xw\cdot\lhd$,
and the head of $A$ is on the first letter of the suffix $w\cdot\lhd$.
Then $A$ looks for the first occurrence from the left of a letter in $w$
that is not translucent for state~$q_1$,
that is, if $w=uav$ such that $u\in(\tau(q_1))^*$ and $a\not\in \tau(q_1)$,
then $A$ nondeterministically chooses a state $q_2\in\delta(q_1,a)$,
erases the letter $a$ from the tape, thus producing the tape contents $xuv\cdot\lhd$, sets
its internal state to~$q_2$, and continues the computation from the configuration $xuq_2v\cdot\lhd$.
In case $\delta(q_1,a)=\emptyset$,
$A$ halts without accepting.
Finally, if $w\in(\tau(q_1))^*$,
then $A$ reaches the end-of-tape marker~$\lhd$ and a transition from the set $\delta(q_1,\lhd)$ is applied.
This transition is either an accept step or a state $q_2$ from~$Q$.
In the former case, $A$ halts and accepts, while in the latter case, it continues
the computation in state $q_2$ by reading its tape again from left to right, that is, from the configuration $q_2xw\cdot\lhd$.
Finally, if $\delta(q_1,\lhd)$ is undefined, then $A$ halts and rejects.
Thus, the computation relation $\vdash_A$ that $A$ induces on its set of configurations
$\Sigma^*\cdot Q\cdot\Sigma^*\cdot\lhd\,\cup\,\{\Accept,\Reject\}$ is the reflexive and transitive closure $\vdash_A^*$
of the single-step computation relation $\vdash_A$ that is specified as follows:
$$xqw\cdot\lhd \vdash_A \left\{ \begin{array}{ll}
     xuq'v\cdot\lhd, & \mbox{if }w=uav,\,u\in(\tau(q))^*,\,a\not\in\tau(q),\mbox{ and }q'\in\delta(q,a),\\
      \Reject,       & \mbox{if }w=uav,\, u\in(\tau(q))^*,\,
      a\not\in\tau(q_0),\mbox{ and }\delta(q,a)=\emptyset,\\
      q'xw\cdot\lhd & \mbox{if }w\in(\tau(q))^*\mbox{ and } q'\in\delta(q,\lhd),\\
     \Accept,      & \mbox{if }w\in(\tau(q))^* \mbox{ and }\delta(q,\lhd)=\Accept,\\
     \Reject,      & \mbox{if }w\in(\tau(q))^*\mbox{ and } \delta(q,\lhd)=\emptyset.
     \end{array}\right.
$$

To describe computations of nrNFAwtls in a compact way, we introduce the notions of a sweep and a cycle.

\begin{definition}\label{DefCycle}
Let $A=(Q,\Sigma,\lhd,\tau,I,\delta)$ be an nrNFAwtl.
\begin{enumerate}
\item[{\rm (a)}] A \emph{sweep} is a part of a computation of $A$ in which the head moves from left to right across the complete
tape contents.
Thus, a sweep has the form
$q_1wu\cdot\lhd\vdash_A^* w'q_2u\cdot\lhd$, where $q_1,q_2\in Q$, $u\in (\tau(q_2))^*$,
the word $w'$ is obtained from $w$ by deleting some letters or $w=w'=\lambda$ and $q_1=q_2$,
and the end-of-tape marker~$\lhd$ is not visited during this partial computation.
We use the notation $$q_1wu\cdot\lhd \vdash_A^s w'uq_2\lhd$$ to denote the above sweep.
Observe that the configurations $w'q_2u\cdot\lhd$ and $w'uq_2\lhd$ have exactly the same immediate successor
configurations, as the word $u$ only contains letters that are translucent for the state~$q_2$.
\item[{\rm (b)}] A \emph{cycle} is a part of a computation of $A$ that consists of a sweep
$q_1wu\cdot\lhd\vdash_A^s w'uq_2\lhd$ together with the next transitional step $q_3\in\delta(q_2,\lhd)$.
Thus, a cycle has the form $q_1wu\cdot\lhd \vdash_A^* w'q_2u\cdot\lhd \vdash_A q_3w'u\cdot\lhd$.
We use the notation $$q_1wu\cdot\lhd \vdash_A^c q_3w'u\cdot\lhd$$ for this cycle.
\end{enumerate}
\end{definition}

A word $w\in\Sigma^*$ is accepted by the nrNFAwtl $A=(Q,\Sigma,\lhd,\tau,I,\delta)$ if there exists
an initial state $q_0\in I$ such that $A$ has an accepting computation of the form $q_0w\cdot\lhd \vdash_A^* \Accept$.
Then
$$L(A) = \{\,w\in\Sigma^*\mid w\mbox{ is accepted by }A\,\}$$
is the \emph{language accepted by}~$A$.
We use $\mathcal{L}({\sf nrNFAwtl})$ to denote the class of languages that are accepted by nrNFAwtls.

\begin{definition}\label{DefnrDFAwtl}
An nrNFAwtl $A= (Q,\Sigma,\lhd,\tau,I,\delta)$ is a
\emph{non-returning deterministic finite-state acceptor with translucent letters},
abbreviated as \emph{nrDFAwtl},
if $|I|=1$ and
if $|\delta(q,a)|\le 1$ for all $q\in Q$ and all $a\in\Sigma\cup\{\lhd\}$.
Then $\mathcal{L}({\sf nrDFAwtl})$ denotes the class of all languages
that are accepted by nrDFAwtls.
\end{definition}

We illustrate these definitions by an example.

\begin{example}\label{Ex1}
Let $A=(Q,\{a,b,c\},\lhd,\tau,\{q_a\},\delta)$ be the nrDFAwtl that is defined by taking $Q=\{q_a,q_b,q_c,q_r\}$,
$\tau(q_a) =\emptyset$, $\tau(q_b) = \{a\}$, $\tau(q_c)=\{b\}$, $\tau(q_r) = \{c\}$, and
$\delta(q_a,a) = q_b$, $\delta(q_b,b)=q_c$, $\delta(q_c,c) = q_r$, $\delta(q_r,\lhd) = q_a$, $\delta(q_a,\lhd) = \Accept$.
Given the word $w=aabbcc$ as input, the automaton $A$ executes the following accepting computation:
$$\begin{array}{lclclclcl}
q_aaabbcc\cdot\lhd & \vdash_A & q_babbcc\cdot\lhd & \vdash_A & aq_cbcc\cdot\lhd & \vdash_A & abq_rc\cdot\lhd\\
                   & \vdash_A & q_aabc\cdot\lhd   & \vdash_A & q_bbc\cdot\lhd    & \vdash_A & q_cc\cdot\lhd \\
                   & \vdash_A & q_r\lhd           & \vdash_A & q_a\lhd           &\vdash_A  & \Accept,
\end{array}$$
that is, $A$ accepts on input $w=aabbcc$.
In fact, $q_aaabbcc\cdot\lhd \vdash_A^sabcq_r\lhd$ is a sweep and $q_aaabbcc\cdot\lhd \vdash_A^c q_aabc\cdot\lhd$
is a cycle of~$A$.
Actually, it is easily seen that $L(A) = \{\,a^nb^nc^n\mid n\ge 0\,\}$.
\hspace*{\fill}$\blacksquare$
\end{example}

Recall from~\cite{otto185} that the language $\{\,a^nb^nc^n\mid n\ge 0\,\}$ is not accepted by any NFAwtl.
\vspace{+2mm}

As defined above, an nrNFAwtl $A=(Q,\Sigma,\lhd,\tau,I,\delta)$ may run into an infinite computation.
Just assume that $q$ is a state of $A$, $w\in (\tau(q))^*$, and $q\in \delta(q,\lhd)$.
Then $qw\cdot\lhd \vdash_A qw\cdot\lhd \vdash_A qw\cdot\lhd$, and so forth.
However, we can avoid this by converting $A$ into an equivalent nrNFAwtl $B$ as follows.
\vspace{+2mm}

Let $B=(Q',\Sigma,\lhd,\tau',I',\delta')$, where $Q' = \{\,(q,S) \mid q\in Q\mbox{ and }S\subseteq Q\,\}$,
$I'=\{\,(q,\emptyset)\mid q\in I\, \}$,
$\tau'(q,S) = \tau(q)$ for all $q\in Q$ and all $S\subseteq Q$,
$\delta'((q,S),a) = \{\,(p,\emptyset) \mid p\in\delta(q,a)\,\}$ for all $q\in Q$, $S\subseteq Q$, and all $a\in\Sigma$,
and $\delta'((q,S),\lhd)  = \{\,(p,S\cup\{q\})\mid p\in\delta(q,\lhd)\mbox{ and } q\not\in S\,\}$ for all $q\in Q$ and all $S\subseteq Q$.
Finally, take $\delta'((q,S),\lhd) = \Accept$ if $\delta(q,\lhd) = \Accept$.
The set $S$ is used to record those states in which the end-of-tape marker has been reached and the computation has continued.
In the next cycle, when a non-translucent letter is read, then this set is emptied, otherwise,
the next state is added to it. This process continues until either a letter is read and deleted, or until
no new state can be added to the current set $S$, in which case the computation fails.
We illustrate this construction through a simple example.

\begin{example}\label{Ex2}
Let $A=(Q,\{a,b\},\lhd,\tau,\{p\},\delta)$, where $Q=\{p,q,r\}$, $\tau(p) = \tau(q)=\tau(r)=\{a\}$,
and $$\delta(p,b) = q, \delta(p,\lhd) = \{q,r\}, \delta(q,\lhd) = p, \delta(r,\lhd) = \Accept,$$
and let $w=aabaa$.
On input $w$, $A$ can execute the following infinite computation:
$$pw\cdot\lhd = paabaa\cdot\lhd \vdash_A aaqaa\cdot\lhd \vdash_A paaaa\cdot\lhd \vdash_A qaaaa\cdot\lhd \vdash_A paaaa\cdot\lhd \vdash_A \cdots$$
The automaton $B=(Q',\{a,b\},\lhd,\tau',\{p\},\delta')$ that is obtained from $A$ through the construction presented above
simulates this computation as follows:
$$(p,\emptyset)aabaa\cdot\lhd \vdash_B aa(q,\emptyset)aa\cdot\lhd \vdash_B (p,\{q\})aaaa\cdot\lhd \vdash_B (q,\{p,q\})aaaa\cdot\lhd \vdash_B \Reject,$$
that is, it recognizes the repetition and aborts the computation.
Of course, using the transition $r\in\delta(p,\lhd)$ or $(r,\{p,q\})\in\delta'((p,\{q\}),\lhd)$, both $A$ and $B$ can accept.
\hspace*{\fill}$\blacksquare$
\end{example}

In general, an nrNFAwtl $A=(Q,\Sigma,\lhd,\tau,I,\delta)$ may accept without having read and deleted its input completely.
This happens for the automata in Example~\ref{Ex2} as $\tau(p)=\tau(r)=\{a\}$, $r\in\delta(p,\lhd)$, and $\delta(r,\lhd)=\Accept$.
However, we can easily convert the nrNFAwtl $A$ into an equivalent nrNFAwtl $C$ that always reads and deletes its input completely before it accepts.
Just take $C=(Q\cup\{q_e\},\Sigma,\lhd,\tau',I,\delta')$, where $q_e$ is a new state,
$\tau'(q)=\tau(q)$ for all $q\in Q$ and $\tau'(q_e)=\emptyset$, and $\delta'$ is defined as follows:
$$\begin{array}{clcll}
- & \delta'(q,a) & = & \delta(q,a) & \mbox{for all }q\in Q\mbox{ and all }a\in\Sigma,\\
- & \delta'(q,\lhd) & = & \multicolumn{2}{l}{\left\{\begin{array}{ll}
                                 \delta(q,\lhd), & \mbox{if }\delta(q,\lhd)\not=\Accept,\\
                                 \{q_e\},        & \mbox{if }\delta(q,\lhd) = \Accept,
                                 \end{array}\right.}\\
- & \delta'(q_e,a) & = & \{q_e\} & \mbox{for all }a\in\Sigma,\\
- & \delta'(q_e,\lhd) & = & \Accept.
\end{array}$$
Given a word $w\in\Sigma^*$ as input, the nrNFAwtl $C$ will execute exactly the same steps as the nrNFAwtl $A$ until $A$ accepts.
Now the accept step of $A$ is simulated by $C$ through changing into state~$q_e$.
As $\tau'(q_e) = \emptyset$ and as $\delta'(q_e,a)=\{q_e\}$ for all $a\in\Sigma$,
$C$ will now read and delete the remaining tape contents and accept on reaching the end-of-tape marker~$\lhd$.
It follows easily that $L(C)=L(A)$.
Together the two constructions considered yield the following technical result.

\begin{proposition}\label{LemNFnrNFAwtl}
Each nrNFAwtl $A$ can effectively be converted into an equivalent nrNFAwtl $C$ that never gets into an infinite computation
and that accepts only after reading and deleting its tape contents completely.
In addition, if $A$ is deterministic, then so is~$C$.
\end{proposition}

\section{Proper Inclusion Results and Incomparability Results}\label{sec2}

First we show that the nrNFAwtl is indeed an extension of the NFAwtl.

\begin{theorem}\label{ThmNFAwtlInnrNFAwtl}
From a given NFAwtl $A$, one can construct an nrNFAwtl $B$ such that $L(B)=L(A)$.
In addition, if $A$ is deterministic, then so is~$B$.
\end{theorem}

\beginproof Let $A = (Q,\Sigma,\lhd,\tau,I,F,\delta)$ be an NFAwtl.
We define a simulating nrNFAwtl $B = (Q_B,\Sigma,\lhd,\tau_B,I_B,\delta_B)$
as follows:
\begin{itemize}
 \item $Q_B = Q \cup \{\,q'\mid q\in Q\,\}$, where for each state $q\in Q$, $q'$ is an additional auxiliary state, and $I_B=I$,
 \item for each state $q\in Q$, $\tau_B(q) = \tau(q)$ and $\tau_B(q')=\Sigma$,
 \item for each state $q\in Q$ and each letter $a\in \Sigma$, $\delta_B(q,a)= \{\,p'\mid p\in \delta(q,a)\,\}$ and $\delta_B(q',a)=\emptyset$.
 Moreover, $\delta_B(q,\lhd) = \Accept$, if $q\in F$, and $\delta_B(q',\lhd) = \{q\}$.
\end{itemize}
It remains to verify that $B$ just simulates the computations of~$A$.

Assume that $qw\cdot\lhd$ is a configuration of~$A$, that is, $q\in Q$ and $w\in\Sigma^*$.
From the definition of the computation relation $\vdash_A$, we see that there are two cases that we must consider.
\begin{itemize}
\item First assume that $w=uav$ for some word $u\in(\tau(q))^*$ and a letter $a\not\in\tau(q)$.
If $p\in\delta(q,a)$, then $qw\cdot\lhd \vdash_A puv\cdot\lhd$ is a possible step of~$A$.
In this case, $B$ can execute the following sequence of steps:
$$qw\cdot\lhd = quav\cdot\lhd \vdash_B up'v\cdot\lhd \vdash_B puv\cdot\lhd.$$
If $\delta(q,a)=\emptyset$, then $A$ halts and rejects.
However, in this case, also $\delta_B(q,a)=\emptyset$, and hence, $B$ halts and rejects as well.

\item If $w\in(\tau(q))^*$,
then $A$ accepts, if $q\in F$, otherwise, it rejects.
In this case, $B$ just acts likewise.
\end{itemize}
Thus, it follows that $L(A) \subseteq L(B)$.
\vspace{+2mm}

Conversely, if $w\in L(B)$, then it is easily verified that each accepting computation of $B$ on input $w$ is just a simulation
of an accepting computation of $A$ on input~$w$.
It follows that $L(B)=L(A)$.

Finally, the above definition of $B$ shows that $B$ is deterministic, if $A$ is.
This completes the proof of Theorem~\ref{ThmNFAwtlInnrNFAwtl}.\myendproof

Together with Example~\ref{Ex1}, this theorem has the following consequence.

\begin{corollary}\label{CorNFAwtlInnrNFAwtl}
$\mathcal{L}({\sf NFAwtl}) \subsetneq \mathcal{L}({\sf nrNFAwtl})$ and $\mathcal{L}({\sf DFAwtl}) \subsetneq \mathcal{L}({\sf nrDFAwtl})$.
\end{corollary}

It is known that all languages accepted by NFAwtls are necessarily semi-linear, that is,
their images with respect to the Parikh mapping are semi-linear subsets of $\N^m$,
where $m$ is the cardinality of the underlying alphabet.
Does a corresponding result also hold for nrNFAwtls?
First we consider this question for the special case of a unary alphabet.

\begin{proposition}\label{PropUnAlph}
A language $L \subseteq \{a\}^*$ is accepted by an nrNFAwtl if and only if it is a regular language.
\end{proposition}

\beginproof
If $L\subseteq \{a\}^*$ is a regular language, then it is accepted by an NFA and therewith
also by an NFAwtl.
Theorem~\ref{ThmNFAwtlInnrNFAwtl} then shows that $L$ is accepted by an nrNFAwtl.
\vspace{+2mm}

Conversely, assume that a language $L\subseteq\{a\}^*$ is accepted by an
nrNFAwtl $A=(Q,\{a\},\lhd,\tau,I,\delta)$.
By Proposition~\ref{LemNFnrNFAwtl}, we can assume that
the nrNFAwtl $A$ never gets into an infinite computation
and that it accepts only after reading and deleting its tape contents completely.
From $A$ we now construct an NFA with $\lambda$-transitions
$B=(Q,\{a\},I,F,\delta_B)$ by taking
$F=\{\,q\in Q\mid \delta(q,\lhd) = \Accept\,\}$
and by defining the transition relation $\delta_B$ as follows:
$$\begin{array}{clcll}
(1) & \delta_B(q,a) & = & \delta(q,a) &\mbox{for all }q\in Q,\\
(2) & \delta_B(q,\lambda) & = & \delta(q,\lhd), &\mbox{if }\tau(q)=\{a\}\mbox{ and }\delta(q,\lhd)\subseteq Q.
\end{array}$$
We claim that $L(B)=L(A)=L$ holds, which then implies that $L$ is a regular language.

For each state $q\in Q$, if $\tau(q)\not=\emptyset$, then $\tau(q) = \{a\}$ and $\delta(q,a)$ is undefined.
Hence,
$$xqw\cdot\lhd \vdash_A \left\{\begin{array}{ll}
                                      q'xw\cdot\lhd, & \mbox{if }q'\in\delta(q,\lhd),\\
                                      \Accept,       & \mbox{if }\delta(q,\lhd) = \Accept,\\
                                      \Reject,       & \mbox{if }\delta(q,\lhd)=\emptyset.
                                     \end{array}\right.$$
On the other hand, for each state $q\in Q$ for which $\tau(q)=\emptyset$,
$$qaw\cdot\lhd \vdash_A \left\{\begin{array}{ll}
                               q'w\cdot\lhd, &\mbox{if }q'\in\delta(q,a),\\
                               \Reject,      &\mbox{if }\delta(q,a)=\emptyset.
                              \end{array}\right.$$
Hence, if $a^m$ is accepted by the nrNFAwtl~$A$,
then a corresponding accepting computation of $A$
reads (and deletes) the word $a^m$ simply letter by letter from left to right,
where this sequence of computational steps may be interspersed with steps
that change the state without reading (and deleting) a letter~$a$.
Now it is easily seen that the NFA $B$ can execute the very same computation.
Conversely, each accepting computation of the NFA $B$ just mirrors an accepting computation
of the nrNFAwtl~$A$.
This completes the proof of Proposition~\ref{PropUnAlph}.\myendproof

Thus, all unary languages that are accepted by nrNFAwtls are semi-linear.
For non-unary alphabets, the corresponding question is still open.
To illustrate this problem, we consider the following detailed example.

\begin{example}\label{ExExp3}
We define the nrDFAwtl $A_{\rm ex3}=(Q,\Sigma,\lhd,\tau,I,\delta)$ as follows:
\begin{itemize}
 \item $Q=\{q_0,q_1,q_2,q_3,q_4,q_5,q_6,q_7,q_8\}$, $\Sigma = \{a,b,c\}$, and $I=\{q_0\}$,
 \item $\begin{array}[t]{cccccccc}
\tau(q_0) & = & \{a\}, & \tau(q_1) & = & \tau(q_2) & = & \emptyset,\\
\tau(q_3) & = & \{b\}, & \tau(q_4) & = & \tau(q_5) & = & \emptyset,\\
\tau(q_6) & = & \{a,c\}, & \tau(q_7) & = & \tau(q_8) & = & \emptyset,
\end{array}$
 \item and the transition function $\delta$ is defined through
 $$\arraycolsep2pt
 \begin{array}{rlclrlclrlclrlcl}
 (1) & \delta(q_0,b) & = & q_1,\quad &
 (4) & \delta(q_2,\lhd) & = & q_7,\quad &
 (7) & \delta(q_5,b) & = & q_6,\quad &
 (10) & \delta(q_7,a) & = & q_8,\\
 (2) & \delta(q_1,c) & = & q_2, &
 (5) & \delta(q_3,c) & = & q_4, &
 (8) & \delta(q_6,b) & = & q_1, &
 (11)& \delta(q_8,\lhd) & = & \Accept.\\
 (3) & \delta(q_2,a) & = & q_3, &
 (6) & \delta(q_4,a) & = & q_5, &
 (9)  & \delta(q_6,\lhd) & = & q_0,\\
\end{array}$$
\end{itemize}

We can actually describe the nrDFAwtl $A_{\rm ex3}$ through the diagram given in Figure~\ref{FigExExp3}.
In this diagram, the vertices correspond to the states of $A$, an edge of the form
{\footnotesize $\xymatrix @R=2pc@C=2pc{
*++[o][F-]{q_i}\ar[r]^{x} & *++[o][F-]{q_j}\\
}$}
denotes a transition from $q_i$ to $q_j$ that simply reads an occurrence of the letter~$x$,
and an edge of the form
{\footnotesize $\xymatrix @R=2pc@C=2.5pc{
*++[o][F-]{q_i}\ar[r]^{(Y^*,x)} & *++[o][F-]{q_j}\\
}$}
denotes a transition from $q_i$ to $q_j$ in which a factor from $Y^*$ is skipped and a subsequent
occurrence of the letter~$x$ is read.
Finally, an edge the label of which contains the end-of-tape marker $\lhd$ corresponds to a restart or an accept operation.

\begin{figure}[ht]
\begin{center}
{\small
$$ \xymatrix @R=2pc@C=3pc{
      &                 & *++[o][F-]{q_6}\ar[dl]_{(\{a,c\}^*,\lhd)}\ar[d]^{(\{a,c\}^*,b)} & *++[o][F-]{q_5}\ar[l]_{b} & *++[o][F-]{q_4}\ar[l]_{a}\\
\ar[r] & *++[o][F-]{q_0}\ar[r]^{(a^*,b)} & *++[o][F-]{q_1}\ar[r]^{c} & *++[o][F-]{q_2}\ar[r]^{a}\ar[d]^{\lhd} & *++[o][F-]{q_3}\ar[u]_{(b^*,c)}\\
       &                 &                 & *++[o][F-]{q_7}\ar[r]^{a} & *++[o][F-]{q_8}\ar[r]^{\lhd} & *++[o][F-]{{\sf Accept}}
} $$
}
\caption{The diagram describing the nrDFAwtl $A_{\rm ex3}$}\label{FigExExp3}
\end{center}
\end{figure}
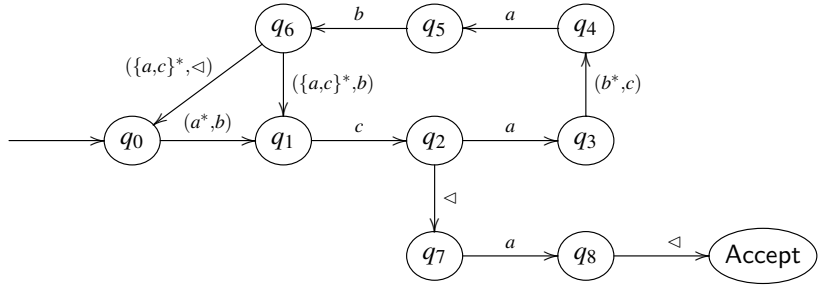

From this diagram, we can easily extract the following information on computations of $A_{\rm ex3}$:
\begin{enumerate}
 \item The shortest path from $q_0$ to {\sf Accept} removes a single occurrence of each of the letters $a$, $b$, and $c$.
 \item A sweep of the automaton $A$ starts in $q_0$ and ends in $q_2$ or in $q_6$, or it starts in~$q_7$.
 \item A sweep may contain one or more repetitions of the cycle $q_6 \to q_1\to q_2\to q_3\to q_4\to q_5\to q_6$,
 during which two occurrences of each of the letters $a$, $b$, and $c$ are removed.
 \item During a sweep that ends at $q_6$, the same even number of occurrences of each of the letters $a$, $b$, and $c$ are removed.
 \item The last part of an accepting computation leads from $q_0$ or from $q_6$ to $q_1$, then to $q_2$,
 then to $q_7$, and on to $q_8$.
 Thus, during this part, a single occurrence of each of the letters $a$, $b$, and $c$ is removed.
\end{enumerate}
Together these observations imply that during each accepting computation, $A_{\rm ex3}$ removes the same uneven number
of occurrences of the letters $a$, $b$, and $c$.
This implies that the Parikh image $\pi(L(A_{\rm ex3}))$ of the language $L(A_{\rm ex3})$ satisfies
the inclusion $\pi(L(A_{\rm ex3}))\subseteq \{\,(2n+1,2n+1,2n+1) \mid n\ge 0\,\}$.
\vspace{+2mm}

We now consider an input of the form $(abc)^{3n}$ for some $n\ge 1$.
This input yields the following computation:
$$\arraycolsep1pt
\begin{array}{lclclcl}
q_0(abc)^{3n}\cdot\lhd & = & q_0abc(abc)^{3n-1}\cdot\lhd & \vdash_{A_{\rm ex3}} & aq_1c(abc)^{3n-1}\cdot\lhd  & \vdash_{A_{\rm ex3}} & aq_2abc(abc)^{3n-2}\cdot\lhd \\
                & \vdash_{A_{\rm ex3}}& aq_3bc(abc)^{3n-2}\cdot\lhd & \vdash_ {A_{\rm ex3}}& abq_4abc(abc)^{3n-3}\cdot\lhd & \vdash_ {A_{\rm ex3}}& abq_5bc(abc)^{3n-3}\cdot\lhd \\
                                   & \vdash_{A_{\rm ex3}}& abq_6cabc(abc)^{3n-4}\cdot\lhd
                                   & \vdash_{A_{\rm ex3}}& abcaq_1c(abc)^{3n-4}\cdot\lhd
                                   & \vdash_{A_{\rm ex3}}& abcaq_2abc(abc)^{3n-5}\cdot\lhd  \\
                                   & \vdash_{A_{\rm ex3}}^* & (abc)^{n-1}aq_2abc(abc)\cdot\lhd
                                   &\vdash_{A_{\rm ex3}} & (abc)^{n-1}aq_3bc(abc)\cdot\lhd
                                   & \vdash_{A_{\rm ex3}} & (abc)^{n-1}abq_4abc\cdot\lhd \\
                                   & \vdash_{A_{\rm ex3}} & (abc)^{n-1}abq_5bc\cdot\lhd
                                   & \vdash_{A_{\rm ex3}} & (abc)^{n-1}abq_6c\cdot\lhd
                                   & \vdash_{A_{\rm ex3}} & q_0(abc)^{n-1}abc\cdot \lhd \\
                                   & = & q_0(abc)^n\cdot\lhd.
\end{array}$$
Finally, as
$$q_0abc\cdot\lhd \vdash_{A_{\rm ex3}} aq_1c\cdot\lhd \vdash_{A_{\rm ex3}} aq_2\lhd \vdash_{A_{\rm ex3}} q_7a\cdot\lhd
\vdash_{A_{\rm ex3}} q_8\lhd \vdash_{A_{\rm ex3}} \Accept,$$
it follows that $L_{\rm exp3} = \{\,(abc)^{3^n}\mid n\ge 0\,\}\subseteq L(A_{\rm ex3})$.
Unfortunately, $A_{\rm ex3}$ also accepts some words that do not belong to the language $L_{\rm exp3}$.
In fact, it can be shown that $L' = \{\,ab(cacabb)^nc\mid n\ge 0\,\}\subseteq L(A_{\rm ex3})$.
Indeed, for $n=0$, we have $ab(cacabb)^nc= abc\in L(A_{\rm ex3})$.
Now, proceeding by induction on $n$,
$$\begin{array}{lclcl}
q_0ab(cacabb)^{n+1}c\cdot\lhd & = & q_0abcacabb(cacabb)^nc\cdot\lhd & \vdash_{A_{\rm ex3}} & aq_1cacabb(cacabb)^nc\cdot\lhd\\
                        & \vdash_{A_{\rm ex3}}^6 & aq_1(cacabb)^nc\cdot\lhd & \vdash_{A_{\rm ex3}}^{*} & \Accept.
\end{array}$$
As $\pi(ab(cacabb)^nc) = (2n+1,2n+1,2n+1)$, we see that
$$\pi(L(A_{\rm ex3})) = \{\,(2n+1,2n+1,2n+1)\mid n\ge 0\,\},$$
which shows that the language $L(A_{\rm ex3})$ is in fact semi-linear.
\vspace{+2mm}

On the other hand, we have the following fact.
\vspace{2mm}

\noindent
{\bf Claim.} $L(A_{\rm ex3}) \cap (abc)^* = L_{\rm exp3}$.
\vspace{+2mm}

\beginproof
If $w=(abc)^{3n+1}$,
then
$$q_0w\cdot\lhd \vdash_{A_{\rm ex3}}^* (abc)^{n-1}abq_6cabc\cdot\lhd \vdash_{A_{\rm ex3}} (abc)^naq_1c\cdot\lhd \vdash_{A_{\rm ex3}} (abc)^naq_2\lhd
\vdash_{A_{\rm ex3}} q_7(abc)^na\cdot\lhd,$$
and from the configuration $q_7(abc)^na\cdot\lhd$, $A_{\rm ex3}$ accepts only if $n=0$.
Analogously, if $w=(abc)^{3n+2}$,
then
$$\begin{array}{lclclcl}
q_0w\cdot\lhd & \vdash_{A_{\rm ex3}} & (abc)^naq_1cabc\cdot\lhd & \vdash_{A_{\rm ex3}} & (abc)^naq_2abc\cdot\lhd
              & \vdash_{A_{\rm ex3}} & (abc)^naq_3bc\cdot\lhd\\
              & \vdash_{A_{\rm ex3}} & (abc)^nabq_4\lhd &\vdash_{A_{\rm ex3}} & \Reject.
\end{array}$$
Hence, the only powers of $abc$ that $A_{\rm ex3}$ accepts are those of the form $(abc)^m$
for which $m$ is a power of three.
\myendproof

Hence, our example shows that the intersection of a language that is accepted by an nrDFAwtl and a regular set
is not necessarily semi-linear.
\hspace*{\fill}$\blacksquare$
\end{example}

At this point, it remains open whether the class
$\mathcal{L}({\sf nrDFAwtl})$ contains any non-unary languages that are not semi-linear.
\vspace{+2mm}

As all rational trace languages are accepted by NFAwtls, Corollary~\ref{CorNFAwtlInnrNFAwtl}
implies that all rational trace languages are accepted by nrNFAwtls.
However, as shown in~\cite{otto206}, the rational trace language
$$L_\vee=\{\,w\in\{a,b\}^*\mid \exists n\ge 0: |w|_a=n\mbox{ and }|w|_b\in\{n,2n\}\,\}$$
is not accepted by any DFAwtl.
Our next result shows that this language is not even accepted by any nrDFAwtl.

\begin{proposition}\label{Prop1GlobalDet}
$L_\vee\not\in\mathcal{L}({\sf nrDFAwtl})$.
\end{proposition}

\beginproof
We prove this result by contradiction.
So assume that $A=(Q,\Sigma,\lhd,\tau,I,\delta)$ is an nrDFAwtl
that accepts the language~$L_\vee$, where $Q=\{q_0,q_1,\ldots,q_{m-1}\}$, $\Sigma=\{a,b\}$, and $I=\{q_0\}$.
\vspace{+2mm}

\noindent
{\bf Claim.} If $q_ia^{r}b^{s}\cdot\lhd \vdash_A^s a^{r-r_1}b^{s-s_1}q_j\lhd$ is a sweep within an accepting computation of
$A$ on input $a^nb^n$ or $a^nb^{2n}$, then $r_1 \le m$ and $s_1\le m$.
\vspace{+2mm}

\beginproof As $L(A)= L_\vee$, and as an nrDFAwtl only deletes letters during its computation,
 we see that $w=a^rb^s$ is converted into $a^{r-r_1}b^{s-s_1}$ for some $0\le r_1\le r$
 and $0\le s_1\le s$.
 Thus, during the above sweep, $A$ first reads (and deletes) $r_1$ copies of the letter $a$ and then
 it reads (and deletes) $s_1$ copies of the letter~$b$.
 If $r_1>m$, then some state of $A$ appears at least twice while the head of $A$ is still
 inside the prefix $a^r$.
 This implies that by using pumping, $A$ can also execute the sweeps of the form
 $$q_ia^{r+\mu \cdot t}b^s\cdot\lhd \vdash_A^s a^{r-r_1}b^{s-s_1}q_j\lhd$$ for all $\mu\ge 1$
 and some value $1\le t\le m$.
 But then, together with $a^nb^n$ or $a^nb^{2n}$,
 $A$ would also  accept the words $a^{n+\mu\cdot t}b^n$ or $a^{n+\mu\cdot t}b^{2n}$,
 a contradiction.
 It follows that $r_1\le m$, and analogously, it can be shown that $s_1\le m$.
 \myendproof

Now let $n > 3m^2$.
Then $a^nb^n\in L_\vee$, and the computation of $A$ on input $a^nb^n$ is accepting.
It consists of a sequence of sweeps and an accept step, that is, we have
$$\begin{array}{lcccccc}
q_0a^nb^n\cdot\lhd & \vdash_A^s & a^{n-r_1}b^{n-s_1}q_{i_1}\cdot\lhd & \vdash_A & q_{j_1}a^{n-r_1}b^{n-s_1}\cdot\lhd\\
 & \vdash_A^s & a^{n-r_1-r_2}b^{n-s_1-s_2}q_{i_2}\cdot\lhd & \vdash_A & q_{j_2}a^{n-r_1-r_2}b^{n-s_1-s_2}\cdot\lhd\\
 & \vdash_A^s & \ldots & \vdash_A^s & a^{n-r_1-r_2-\cdots -r_k}b^{n-s_1-s_2-\cdots -s_k}q_{i_k}\cdot\lhd\\
 & \vdash_A & \Accept,
\end{array}$$
where $k\ge 1$ and $r_i,s_i\le m$ for all $i=1,2,\ldots,k$.
If $n > r_1+r_2+\cdots +r_k$, then $A$ would also accept the word $a^{n+1}b^n\not\in L_\vee$,
and if $n>s_1+s_2+\cdots +s_k$, then $A$ would also accept the word $a^{n}b^{n+1}\not\in L_\vee$.
It follows that $n=r_1+r_2+\cdots +r_k = s_1+s_2+\cdots +s_k$, that is,
$A$ erases its input $a^nb^n$ completely before it accepts.
Because of the above claim, this means in particular that the number of sweeps~$k$ in the above computation
satisfies the inequality $k>3m$.
\vspace{+2mm}

As $A$ has only $m$ states, it follows that there are indices $1\le \alpha < \beta \le m+1$
such that the states $q_{j_\alpha}$ and $q_{j_\beta}$ are identical.
Hence, the above computation can be written as follows:
$$\begin{array}{lcl}
q_0a^nb^n\cdot\lhd & \vdash_A^* & q_{j_\alpha}a^{n-r_1-r_2-\cdots-r_\alpha}b^{n-s_1-s_2-\cdots-s_\alpha}\cdot\lhd\\
 & \vdash_A^* & q_{j_\beta}a^{n-r_1-r_2-\cdots-r_\alpha-r_{\alpha+1}-\cdots-r_\beta}
 b^{n-s_1-s_2-\cdots-s_\alpha-s_{\alpha+1}-\cdots-s_\beta}\cdot\lhd\\
 & = & q_{j_\alpha}a^{n-r_1-r_2-\cdots-r_\alpha-r_{\alpha+1}-\cdots-r_\beta}
 b^{n-s_1-s_2-\cdots-s_\alpha-s_{\alpha+1}-\cdots-s_\beta}\cdot\lhd\\
 & \vdash_A^* & q_{i_k}\lhd \vdash_A \Accept.
\end{array}$$
To simplify the notation, we take 
$n_\alpha  =  n-r_1-r_2-\cdots-r_\alpha$,
$c  =  r_{\alpha+1}+\cdots+r_\beta$,
$c'  =  s_{\alpha+1}+\cdots+s_\beta$, and
$n_\alpha'  =  n-s_1-s_2-\cdots-s_\alpha$. 
Then we also have the following accepting computation:
$$\arraycolsep2pt
\begin{array}{lcllclclc}
q_0a^{n+c}b^{n+c'}\cdot\lhd & \vdash_A^* & q_{j_\alpha}a^{n_\alpha+c}b^{n'_\alpha+c'}\cdot\lhd
     & \vdash_A^* & q_{j_\alpha}a^{n_\alpha-c+c}b^{n'_\alpha-c'+c'}\cdot\lhd
     & = & q_{j_\alpha}a^{n_\alpha}b^{n'_\alpha}\cdot\lhd
     & \vdash_A^*& \Accept.
\end{array}$$
Thus, $a^{n+c}b^{n+c'}\in L_\vee$.
As $c\le m^2$ and $c'\le m^2$, while $n > 3m^2$,
it follows that $n+c=n+c'$, which in turn implies that $c=c'$.
\vspace{+2mm}

Now we consider the accepting computation of $A$ for the input $a^{n}b^{2n}\in L_\vee$.
As $A$ is deterministic, this computation looks as follows:
$$\arraycolsep2pt
\begin{array}{lcccccc}
q_0a^nb^{2n}\cdot\lhd & \vdash_A^* & q_{j_\alpha}a^{n_\alpha}b^{n+n'_\alpha}\cdot\lhd
                      & \vdash_A^* & q_{j_\alpha}a^{n_\alpha-c}b^{n+n'_\alpha-c}\cdot\lhd
                      & \vdash_A^* & \Accept.
 \end{array}$$
However, $A$ can then also execute the following accepting computation:
$$\arraycolsep2pt
\begin{array}{lcccccccc}
q_0a^{n+c}b^{2n+c}\cdot\lhd & \vdash_A^* & q_{j_\alpha}a^{n_\alpha+c}b^{n+n'_\alpha+c}\cdot\lhd
  & \vdash_A^* & q_{j_\alpha}a^{n_\alpha-c+c}b^{n+n'_\alpha-c+c}\cdot\lhd
  &  =         & q_{j_\alpha}a^{n_\alpha}b^{n+n'_\alpha}\cdot\lhd
  & \vdash_A^* & \Accept.
\end{array}$$
Thus, $a^{n+c}b^{2n+c}\in L_\vee$.
However, $n+c<2n+c < 2(n+c)$, which means that $a^{n+c}b^{2n+c}\not\in L_\vee$, a contradiction.
This shows that $L_\vee$ is not accepted by any nrDFAwtl.
\myendproof

\vspace{-4mm}

It thus follows that the class of rational trace languages is not contained in the language
class $\mathcal{L}({\sf nrDFAwtl})$.
Finally, we consider the subset $P_2$ of the semi-Dyck language~$D_2$ that is defined by the context-free grammar
$$G=(\{S\},\{a,b,c,d\},S,\{(S\to\lambda),(S\to aSc),(S\to bSd)\}.$$
Thus, $P_2$ consists of all fully bracketed expressions over $\Sigma=\{a,b,c,d\}$,
where the letters $a$ and $b$ are seen as opening brackets, the letters $c$ and $d$ are
the corresponding closing brackets, and all opening brackets come before all closing brackets..
Hence, $P_2$ consists of all words of the form
$$w=a^{i_1}b^{j_1}a^{i_2}b^{j_2}\cdots a^{i_k}b^{j_k}d^{j_k}c^{i_k}\cdots d^{j_2}c^{i_2}d^{j_1}c^{i_1},$$
where $k\ge 0$, $i_1,j_k\ge 0$, $i_2,i_3,\ldots,i_k>0$, and $j_1,j_2,\ldots,j_{k-1}>0$.
It is easily seen that
$P_2$ is a deterministic linear language, that is,
it is accepted by a deterministic one-turn pushdown automaton (see, e.g.,~\cite{ABB97}).
For this language we have the following negative result.

\begin{theorem}\label{ThmP2NotInNRNFAwtl}
$P_2\not\in\mathcal{L}({\sf nrNFAwtl})$.
\end{theorem}

\noindent
{\bf Proof outline.} We prove this result by contradiction.
So assume that $A=(Q,\Sigma,\lhd,\tau,I,\delta)$ is an nrNFAwtl such that $L(A)=P_2$,
where $Q=\{q_0,q_1,\ldots,q_{m-1}\}$ and $\Sigma=\{a,b,c,d\}$.

Let $w\in\Sigma^*$ be an input word.
A factor of $w$ of maximum length that only consists of occurrences of the letter~$a$ is called an \emph{$a$-block} of $w$,
and analogously, we have \emph{$b$-blocks}, \emph{$c$-blocks}, and \emph{$d$-blocks}.
As observed above, $w$ consists of an alternating sequence of $a$- and $b$-blocks that is followed by an alternating sequence of $c$- and $d$-blocks.
It is important to notice that the language $P_2$ contains words that consist of arbitrarily many blocks of arbitrary size.
Based on this observation the following technical results
concerning accepting computations of $A$ on inputs of sufficient size can be derived:
\begin{enumerate}
 \item[(1)] During a sweep within an accepting computation of $A$ on an input $w$,
 at most $m$ letters can be deleted from any block of size larger than~$m$.
 \item[(2)] During a sweep within an accepting computation on an input~$w$, $A$ deletes letters from adjacent blocks from the
$\{a,b\}^*$-prefix of $w$ starting with the first $a$-block or the first $b$-block, and analogously,
it deletes letters from adjacent blocks of the $\{c,d\}^*$-suffix,
starting with the first $d$-block or the first $c$-block.
\item[(3)] Within an accepting computation of $A$,
the number of blocks that are modified within a given sweep
is bounded from above by a fixed multiple of the number of sweeps already executed.
\item[(4)] For any word of the form
$w=a^{i_1}b^{j_1}\cdots a^{i_k}b^{j_k}d^{j_k}c^{i_k}\cdots d^{j_1}c^{i_1} \in P_2,$
where $k>6m^2+2$ and $i_\nu,j_\nu>m^2$ for all $\nu=1,2,\ldots,k$,
any accepting computation of $A$ on input $w$ consists of at most $m$ cycles.
\end{enumerate}
These statements
are proved in a similar way as Proposition~\ref{Prop1GlobalDet}.
However, more involved arguments based on pumping are needed.
Based on these statements,
we can now complete the proof as follows.

If $w\in P_2$ is a word of the form described in (4),
then an accepting computation of $A$ on input $w$ consists of at most $m$ cycles.
During each cycle, at most $m$ letters are deleted from at most $4m\cdot (m+1) = 4m^2+4m$
blocks, and so,
the word obtained through these cycles still consists of $4k > 24m^2+8$ non-empty blocks.
As $A$ now accepts, it is obvious that together with the word~$w$,
$A$ also accepts words that do not belong to the language~$P_2$.
This contradiction shows that there is no nrNFAwtl $A$ such that $L(A)=P_2$.
This completes the proof of Theorem~\ref{ThmP2NotInNRNFAwtl}.
\myendproof

\begin{figure}[t]
{\small
\begin{center}
\mbox{
$\xymatrix@R10pt@C6pt{
\txt{\phantom{\sf DCFL}} & \txt{\sf CSL}\\
 &  \txt{\sf GCSL} \ar@{->}[u]
  & \txt{$\mathcal{L}(\mbox{\sf nrNFAwtl})$}\ar@{->}[ul]\\
 & \txt{\sf CFL}\ar@{->}[u]  &  & \txt{$\mathcal{L}(\mbox{\sf nrDFAwtl})$}\ar@{->}[ul] \\
 & \txt{\sf LIN}\ar@{->}[u] & \txt{$\mathcal{L}(\mbox{\sf NFAwtl})$}\ar@{->}[uu] & \\
 & \txt{\sf DLIN}\ar@{->}[u] &  \txt{$\sf{LRAT}$} \ar@{->}[u]
 & \txt{$\mathcal{L}(\mbox{\sf DFAwtl})$}\ar@{->}[uu]\ar@{->}[lu]\\
& \txt{\sf REG}\ar@{->}[u] \ar@{->}[ru] & \txt{$\mathcal{L}({\sf NFA})$}\ar@{=}[l]\ar@{=}[r] & \txt{$\mathcal{L}({\sf DFA})$}\ar@{->}[u]\\
}$
¸}
\caption{Hierarchy of language classes accepted by the various types
of finite automata with translucent letters.
Here $\sf{LRAT}$ denotes the class of all rational trace languages.
Each arrow represents a proper inclusion, and classes that are not
connected by a sequence of arrows are incomparable under inclusion.}\label{Fig1}
\end{center}
}
\end{figure}
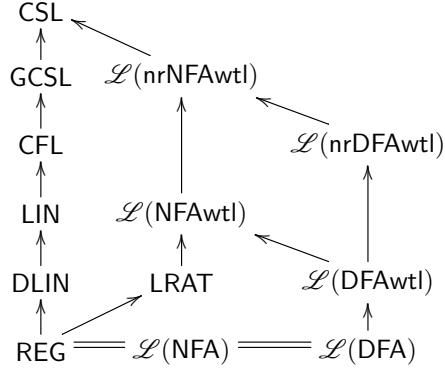

Thus, the nrNFAwtl does not even accept all deterministic linear languages.
The diagram in Figure~\ref{Fig1} summarizes the relationships between the classes
of languages that are accepted by the various types of finite automata with
translucent letters and the classes of the Chomsky hierarchy.

\section{Closure and Non-Closure Properties}\label{sec3}
Here we present some closure and non-closure properties for the classes of languages
that are accepted by nrNFAwtls and by nrDFAwtls.

\begin{theorem}\label{ThmClosureNRNFAwtl}
The language class $\mathcal{L}({\sf nrNFAwtl})$ is closed under union and disjoint shuffle.
\end{theorem}

\beginproof
Let $A_1=(Q_1,\Sigma,\lhd,\tau_1,I_1,\delta_1)$ and $A_2=(Q_2,\Sigma,\lhd,\tau_2,I_2,\delta_2)$
be two nrNFAwtls.
Without loss of generality we may assume that the sets $Q_1$ and $Q_2$ are disjoint.
Let $A=(Q_1\cup Q_2,\Sigma,\lhd,\tau,I_1\cup I_2,\delta)$ be the nrNFAwtl that is defined by taking
$$\tau(q) = \left\{\begin{array}{ll}
                         \tau_1(q), & \mbox{if }q\in Q_1\\
                         \tau_2(q), &\mbox{if }q\in Q_2\\
                        \end{array}\right\}
\mbox{ and }
\delta(q,a) = \left\{\begin{array}{ll}
                           \delta_1(q,a), & \mbox{if }q\in Q_1\\
                           \delta_2(q,a), &\mbox{if }q\in Q_2
                          \end{array}\right\}$$
                          for all $a\in\Sigma\cup\{\lhd\}$.
Then $L(A) = L(A_1)\cup L(A_2)$, which proves that the class $\mathcal{L}({\sf nrNFAwtl})$
is closed under union.
\vspace{+2mm}

Let $L_1\subseteq\Sigma_1^*$ and $L_2\subseteq \Sigma_2^*$, where the alphabets $\Sigma_1$ and $\Sigma_2$ are disjoint.
If $A_1=(Q_1,\Sigma_1,\lhd,\tau_1,I_1,\delta_1)$ and $A_2=(Q_2,\Sigma_2,\lhd,\tau_2,I_2,\delta_2)$ are nrNFAwtls
with disjoint sets of states
such that $L(A_1)=L_1$ and $L(A_2) =L_2$,
then we obtain an nrNFAwtl $A=(Q_1\cup Q_2,\Sigma_1\cup\Sigma_2,\lhd,\tau,I_1,\delta)$ for the
shuffle ${\rm sh}(L_1,L_2)$ by taking
$$\tau(q) = \left\{\begin{array}{ll}
                   \tau_1(q)\cup \Sigma_2, & \mbox{if }q\in Q_1\\
                   \tau_2(q)\cup\Sigma_1,  & \mbox{if }q\in Q_2
                   \end{array}\right\},
\delta(q,a) = \left\{\begin{array}{ll}
                    \delta_1(q,a), &\mbox{if }q\in Q_1\mbox{ and }a\in\Sigma_1\\
                    \delta_2(q,a), &\mbox{if }q\in Q_2\mbox{ and }a\in\Sigma_2
                     \end{array}\right\},$$
and $\delta(q,\lhd) = \left\{\begin{array}{ll}
                              \delta_1(q,\lhd), &\mbox{if }q\in Q_1\mbox{ and }\delta_1(q,\lhd)\subseteq Q_1,\\
                              I_2, & \mbox{if }q\in Q_1\mbox{ and }\delta_1(q,\lhd)=\Accept,\\
                              \delta_2(q,\lhd), &\mbox{if }q\in Q_2.
                             \end{array}\right.$
\vspace{+2mm}

Given a word $w \in (\Sigma_1\cup\Sigma_2)^*$ as input,
$A$ starts in a state from $I_1$ and it behaves just like the automaton~$A_1$, ignoring all letters from~$\Sigma_2$.
If and when the end-of-tape marker $\lhd$ is reached in a state $q\in Q_1$ for which $\delta_1(q,\lhd)=\Accept$,
then $A$ enters a state from the set $I_2$ and continues its computation by simulating $A_2$, this time ignoring
all letters from $\Sigma_1$ that may still be on its tape.
Finally, $A$ accepts if and when the computation of $A_2$ accepts.
It follows that $L(A) = {\rm sh}(L(A_1),L(A_2)) = {\rm sh}(L_1,L_2)$.
Thus, the class $\mathcal{L}({\sf nrNFAwtl})$ is closed under disjoint shuffle.
\myendproof

For the nrDFAwtl, we have the following results.

\begin{theorem}\label{ThmClosureNRDFAwtl}
The language class $\mathcal{L}({\sf nrDFAwtl})$ is closed under complementation and disjoint shuffle,
but it is neither closed
under union nor under intersection.
Moreover, this class is not closed under alphabetic morphisms.
\end{theorem}

\beginproof
The language $L_1=\{\,w\in\{a,b\}^*\mid |w|_a=|w|_b\,\}$
and the language $L_2 = \{\,w\in\{a,b\}^*\mid |w|_b = 2\cdot|w|_a\,\}$
are accepted by DFAwtls.
However, $L_1\cup L_2 = L_\vee$, which is not even accepted by any nrDFAwtl by Proposition~\ref{Prop1GlobalDet}.
This shows that the class $\mathcal{L}({\sf nrDFAwtl})$ is not closed under union.
\vspace{+2mm}

Next we prove that the class $\mathcal{L}({\sf nrDFAwtl})$ is closed under complementation.
Let $A=(Q,\Sigma,\lhd,\tau,I,\delta)$ be an nrDFAwtl.
We define an nrDFAwtl $A^c = (Q\cup\{q_+\},\Sigma,\lhd,\tau^c,I,\delta^c)$, where $q_+$ is a new state, by taking
$$\begin{array}{lcl}
\tau^c(q) & = & \left\{\begin{array}{ll}
                     \tau(q), & \mbox{if } q\in Q\\
                     \Sigma,  & \mbox{if }q=q_+
                     \end{array}\right\},
                     \end{array}$$
and by defining, for all $q\in Q$ and all $a\in\Sigma\,\cup\,\{\lhd\}$,
$$\begin{array}{lcl}
\delta^c(q,a) & = & \left\{\begin{array}{ll}
                        \delta(q,a), & \mbox{if }\delta(q,a)\in Q\\
                        q_+, & \mbox{if }a\not\in\tau(q) \mbox{ and }\delta(q,a) \mbox{ is undefined}\\
                        \emptyset, & \mbox{if }a=\lhd \mbox{ and }\delta(q,\lhd) = \Accept
                \end{array}\right\},\mbox{ and}\\
\delta^c(q_+,\lhd) & = &\Accept.
\end{array}$$
Given a word $w\in\Sigma^*$ as input,
the automaton $A^c$ simulates the computation of the automaton $A$ on input $w$ step by step
until $A$ either accepts or gets stuck.
In the former case, $A^c$ reaches the end-of-tape marker $\lhd$ and gets stuck,
while in the latter case it enters the state $q_+$ and accepts.
It follows that $L(A^c) = \Sigma^*\smallsetminus L(A)$, which shows that
the class $\mathcal{L}({\sf nrDFAwtl})$ is closed under complementation.
\vspace{+2mm}

Closure under complementation and non-closure under union
imply that the class $\mathcal{L}({\sf nrDFAwtl})$ is not closed under intersection.
Furthermore, closure under disjoint shuffle is proved in the same way as for nrNFAwtls.
\vspace{+2mm}

Finally, let $\Sigma=\{a,b,c\}$, let
$$L=\{\,w\in\{a,b\}^*\mid |w|_a=|w|_b\,\}\cup\{\,w\in\{a,c\}^*\mid |w|_c=2\cdot|w|_a\,\},$$
and let $\varphi:\Sigma^*\to\{a,b\}^*$ be the alphabetic morphism that is defined through
$a\mapsto a$, $b\mapsto b$, and $c\mapsto b$.
It is easily verified that the language $L$ is accepted by a DFAwtl.
However, $\varphi(L)=L_\vee$, which is not accepted by any nrDFAwtl by Proposition~\ref{Prop1GlobalDet}.
This proves that the class $\mathcal{L}({\sf nrDFAwtl})$ is not closed under alphabetic morphisms.
\myendproof

\section{Decision Problems}\label{sec4}

The membership problem for each nrNFAwtl is solvable in linear space.
Moreover, it is straightforward to see that $\mathcal{L}({\sf nrNFAwtl}) \subseteq {\sf NTIME}(n^2)$.
In particular, the membership problem for a nrDFAwtl is decidable in quadratic time.
However, by associating, for each letter $a\in\Sigma$, a balanced binary search tree
(see, e.g.,~\cite{Co22})
$T_a$ to a word of length $n$ over $\Sigma$ such that $T_a$ contains those indices $i\in\{1,2,\ldots,|w|\}$
at which the letter $a$ occurs in~the word~$w$,
it can be shown that the computation of a nrDFAwtl $A$ on a word of length $n$ can be simulated by a random access machine (a RAM)
in $O(n\cdot \log n)$ steps.
As each operation involves only $\log n$ many bits,
we obtain the following result.

\begin{theorem}\label{ThmCompl}
The membership problem for a nrDFAwtl is decidable in time ${\mathrm O}(n\cdot (\log n)^2)$.
\end{theorem}

The construction shows that
each operation of a non-returning NFAwtl can be simulated nondeterministically by a RAM in $\log n$ many steps.
Hence, we obtain the following obvious corollary.

\begin{corollary}\label{CorCompl}
The membership problem for an nrNFAwtl is in ${\sf NTIME}(n\cdot (\log n)^2)$.
\end{corollary}

In fact, by using an extension of the technique presented by Nagy and Kov{{\'a}}cs in~\cite{NagKov14},
it can be shown that the membership problem for a nrDFAwtl is even decidable in time $O(n\cdot \log n)$
if the underlying alphabet is only of cardinality two.
However, it remains open whether a corresponding result can also be obtained for the case of larger alphabets.

By Proposition~\ref{PropUnAlph}, a unary language is accepted by an nrNFAwtl if and only if it is a regular language.
This implies immediately that the emptiness problem (and the finiteness problem) is decidable
for nrNFAwtls that accept unary languages.
Concerning non-unary languages, the situation is more complicated.

From the diagram describing a given nrNFAwtl $A$, we can immediately extract information on the patterns of the words
that $A$ can scan during a single sweep (or cycle).
Of course, if there is a sweep that starts in an initial state and that reaches a state in which $A$ accepts at the end-of-tape marker,
then the corresponding words are accepted by $A$, which means that they are witnesses for the fact that the language $L(A)$
is non-empty.
In general, however, $A$ may not have any accepting computations that just consist of single sweeps.
In this case, each accepting computation consists of a sequence of sweeps.
Now the words that $A$ scans during these sweeps form a sequence that can be combined into an accepted word.
However, it is not clear whether this can always be done.
To illustrate this problem, we consider a simple example.

\begin{example}\label{ExEmptiness}
Let $A=(Q,\Sigma,\lhd,\tau,q_0,\delta)$ be the nrDFAwtl that is defined by taking
$Q=\{q_0,q_1,q_2,q_3,q_4,q_5\}$, $\Sigma = \{a,b,c\}$, and by defining the functions $\tau$ and $\delta$ as follows:
$$\begin{array}{lcllcllcl}
\tau(q_0) & = & \{b\}, \quad & \tau(q_2) & = & \{c\},\quad & \tau(q_4) & = & \{a\},\\
\tau(q_1) & = & \{c\},       & \tau(q_3) & = & \emptyset, & \tau(q_5) & = & \emptyset,\\[+1mm]
\delta(q_0,a) & = & q_1, &
\delta(q_2,b) & = & q_3, &
\delta(q_4,c) & = & q_5,\\
\delta(q_1,\lhd) & = & q_2, &
\delta(q_3,\lhd) & = & q_4, &
\delta(q_5,\lhd) & = & \Accept.
\end{array}$$
This nrDFAwtl is depicted by the diagram in Figure~\ref{FigExEmptiness}.

\begin{figure}[ht]
\begin{center}
{\small
\[ \xymatrix @R=2pc@C=3pc{
\ar[r] & *++[o][F-]{q_0}\ar[r]^{(b^*,a)} & *++[o][F-]{q_1}\ar[r]^{(c^*,\lhd)} & *++[o][F-]{q_2}\ar[r]^{(c^*,b)} & *++[o][F-]{q_3}\ar[r]^{\lhd}
       & *++[o][F-]{q_4}\ar[r]^{(a^*,c)} & *++[o][F-]{q_5}\ar[r]^{\lhd} & {{\sf Accept}}\\
} \]
}
\end{center}
\caption{The nrDFAwtl $A$ from Example~\ref{ExEmptiness}}\label{FigExEmptiness}
\end{figure}
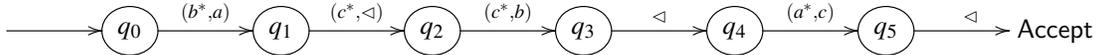

From this diagram, we can immediately extract three sweeps (or rather cycles):
$$\begin{array}{clclcl}
(1) & q_0b^*ac^*\cdot\lhd & \vdash_A & b^*q_1c^*\cdot\lhd & \vdash_A & q_2b^*c^*\cdot\lhd,\\
(2) & q_2c^*b\cdot\lhd    & \vdash_A & c^*q_3\lhd & \vdash_A & q_4c^*\cdot\lhd,\\
(3) & q_4a^*c\cdot\lhd    & \vdash_A & a^*q_5\lhd & \vdash_A & \Accept.
\end{array}$$
However, these three sweeps (or cycles) cannot be combined into an accepting computation of $A$.
The third sweep requires that there is an occurrence of the letter $c$ which may only be preceded by occurrences of the letter~$a$.
The second sweep requires that there is an occurrence of the letter $b$, which may only be preceded by occurrences of the letter~$c$.
Finally, the first sweep requires that  there is an occurrence of the letter $a$ that may only be preceded by occurrences of the letter $b$
and that may only be followed by an occurrence of the letter~$c$.
Together these requirements imply that there is no word that $A$ accepts, that is, $L(A)=\emptyset$.
\hspace*{\fill}$\blacksquare$
\end{example}

Thus, it remains to determine whether, from a given finite set of
patterns of words that are accepted by an nrNFAwtl $A$ in different sweeps (or cycles),
one can extract sufficient information for deciding whether there exists a word that is accepted by $A$,
that is, whether from the various patterns a word can be obtained that is compatible with all these patterns.

\section{Conclusion}\label{sec5}
We have extended the NFAwtl and its deterministic variant, the DFAwtl, to the non-returning NFAwtl and the non-returning DFAwtl
by abandoning the requirement that, in each step, the automaton reads and deletes the first letter from the beginning
of the current word on its tape that is not translucent for the current state.
The non-returning types of automata are indeed more expressive than the original types.
In fact, we presented a complete classification of the resulting language classes in relation to the Chomsky hierarchy.
Also we derived some closure and non-closure properties for these language classes and proved that the membership problem
for a non-returning DFAwtl is of time complexity $O(n\cdot (\log n)^2)$.

However, many questions concerning the nrNFAwtl and the nrDFAwtl are still open.
Here we stress only four of them.

\noindent
\begin{enumerate}
\item Are all languages accepted by nrNFAwtls necessarily semi-linear?
While for unary languages this is indeed the case,
the question remains open for non-unary languages.
\item Is the language class $\mathcal{L}({\sf nrNFAwtl})$ closed under intersection with regular sets?
If it is, then we see from Example~\ref{ExExp3} that $\mathcal{L}({\sf nrNFAwtl})$ contains
languages that are not semi-linear.
However,  we conjecture that $\mathcal{L}({\sf nrNFAwtl})$ is not closed under this operation, as
we expect that the language $L_{\rm exp3}$ is not accepted by any nrNFAwtl.
\item Can the upper bound of $O(n\cdot (\log n)^2)$ for the time complexity of the membership problem for a nrDFAwtl
be improved to $O(n\cdot \log n)$ also for alphabets of cardinality larger than two?
\item Is emptiness decidable for nrNFAwtls or for nrDFAwtls?
\end{enumerate}

Currently, pumping techniques as used in the proofs of Proposition~\ref{Prop1GlobalDet}
and Theorem~\ref{ThmP2NotInNRNFAwtl} are our only means for proving that a given language
is not accepted by any nrNFAwtl.
In order to solve the open problems above, it appears to be necessary to develop other techniques
for this task.

\bibliographystyle{eptcs.bst}
\bibliography{NCMA2022}

\begin{thebibliography}{10}
\providecommand{\bibitemdeclare}[2]{}
\providecommand{\surnamestart}{}
\providecommand{\surnameend}{}
\providecommand{\urlprefix}{Available at }
\providecommand{\url}[1]{\texttt{#1}}
\providecommand{\href}[2]{\texttt{#2}}
\providecommand{\urlalt}[2]{\href{#1}{#2}}
\providecommand{\doi}[1]{doi:\urlalt{https://doi.org/#1}{#1}}
\providecommand{\eprint}[1]{arXiv:\urlalt{https://arxiv.org/abs/#1}{#1}}
\providecommand{\bibinfo}[2]{#2}

\bibitemdeclare{incollection}{ABB97}
\bibitem{ABB97}
\bibinfo{author}{J.-M.\ \surnamestart Autebert\surnameend},
  \bibinfo{author}{J.\ \surnamestart Berstel\surnameend} \&
  \bibinfo{author}{L.~\surnamestart Boasson\surnameend} (\bibinfo{year}{1997}):
  \emph{\bibinfo{title}{Context-free languages and pushdown automata}}.
\newblock In \bibinfo{editor}{G.\ \surnamestart Rozenberg\surnameend} \&
  \bibinfo{editor}{A.~\surnamestart Salomaa\surnameend}, editors: {\slshape
  \bibinfo{booktitle}{Handbook of {F}ormal {L}anguages}}, \bibinfo{volume}{1},
  \bibinfo{publisher}{Springer}, \bibinfo{address}{Berlin, Heidelberg}, pp.
  \bibinfo{pages}{111--174}, \doi{10.1007/978-3-642-59136-5_3}.

\bibitemdeclare{article}{otto84}
\bibitem{otto84}
\bibinfo{author}{G.\ \surnamestart Buntrock\surnameend} \&
  \bibinfo{author}{F.~\surnamestart Otto\surnameend} (\bibinfo{year}{1998}):
  \emph{\bibinfo{title}{Growing context-sensitive languages and
  {C}hurch-{R}osser languages}}.
\newblock {\slshape \bibinfo{journal}{\ICOM}} \bibinfo{volume}{141}, pp.
  \bibinfo{pages}{1--36}, \doi{10.1006/inco.1997.2681}.

\bibitemdeclare{book}{Co22}
\bibitem{Co22}
\bibinfo{author}{T.H.\ \surnamestart Cormen\surnameend}, \bibinfo{author}{C.E.\
  \surnamestart Leiserson\surnameend}, \bibinfo{author}{R.L.\ \surnamestart
  Rivest\surnameend} \& \bibinfo{author}{C.~\surnamestart Stein\surnameend}
  (\bibinfo{year}{2022}): \emph{\bibinfo{title}{Introduction to {A}lgorithms}},
  \bibinfo{edition}{4th} edition.
\newblock \bibinfo{publisher}{{MIT} {P}ress}.

\bibitemdeclare{incollection}{FPS2012}
\bibitem{FPS2012}
\bibinfo{author}{H.\ \surnamestart Fernau\surnameend}, \bibinfo{author}{M.\
  \surnamestart Paramasivan\surnameend} \& \bibinfo{author}{M.L. \surnamestart
  Schmid\surnameend} (\bibinfo{year}{2012}): \emph{\bibinfo{title}{Jumping
  finite automata: {C}haracterizations and complexity}}.
\newblock In \bibinfo{editor}{F.~\surnamestart Drewes\surnameend}, editor:
  {\slshape \bibinfo{booktitle}{{CIAA 2012}, Proc.}}, {\slshape
  \bibinfo{series}{Lecture Notes in Computer Science}} \bibinfo{volume}{9223},
  \bibinfo{publisher}{Springer}, \bibinfo{address}{Heidelberg}, pp.
  \bibinfo{pages}{89--101}, \doi{10.1007/978-3-319-22360-5_8}.

\bibitemdeclare{incollection}{JMPV95}
\bibitem{JMPV95}
\bibinfo{author}{P.\ \surnamestart Jan{{\v c}}ar\surnameend},
  \bibinfo{author}{F.\ \surnamestart Mr{\'a}z\surnameend}, \bibinfo{author}{M.\
  \surnamestart Pl{\'a}tek\surnameend} \& \bibinfo{author}{J.~\surnamestart
  Vogel\surnameend} (\bibinfo{year}{1995}): \emph{\bibinfo{title}{Restarting
  automata}}.
\newblock In \bibinfo{editor}{H.~\surnamestart Reichel\surnameend}, editor:
  {\slshape \bibinfo{booktitle}{{FCT'95}, {P}roc.}}, {\slshape
  \bibinfo{series}{Lecture Notes in Computer Science}} \bibinfo{volume}{965},
  \bibinfo{publisher}{Springer}, \bibinfo{address}{Berlin}, pp.
  \bibinfo{pages}{283--292}, \doi{10.1007/3-540-60249-6_60}.

\bibitemdeclare{incollection}{Lou2007}
\bibitem{Lou2007}
\bibinfo{author}{R.~\surnamestart Loukanova\surnameend} (\bibinfo{year}{2007}):
  \emph{\bibinfo{title}{Linear context free languages}}.
\newblock In \bibinfo{editor}{C.B.\ \surnamestart Jones\surnameend},
  \bibinfo{editor}{Z.\ \surnamestart Liu\surnameend} \&
  \bibinfo{editor}{J.~\surnamestart Woodcock\surnameend}, editors: {\slshape
  \bibinfo{booktitle}{{ICTAC 2007}, Proc.}}, {\slshape \bibinfo{series}{Lecture
  Notes in Computer Science}} \bibinfo{volume}{4711},
  \bibinfo{publisher}{Springer}, \bibinfo{address}{Heidelberg}, pp.
  \bibinfo{pages}{351--365}, \doi{10.1007/978-3-540-75292-9_24}.

\bibitemdeclare{article}{MeZe2012}
\bibitem{MeZe2012}
\bibinfo{author}{A.\ \surnamestart Meduna\surnameend} \&
  \bibinfo{author}{P.~\surnamestart Zemek\surnameend} (\bibinfo{year}{2012}):
  \emph{\bibinfo{title}{Jumping finite automata}}.
\newblock {\slshape \bibinfo{journal}{\IJFCS}} \bibinfo{volume}{23}, pp.
  \bibinfo{pages}{1555--1578}, \doi{10.1142/S0129054112500244}.

\bibitemdeclare{article}{Mra01}
\bibitem{Mra01}
\bibinfo{author}{F.~\surnamestart Mr{{\'a}}z\surnameend}
  (\bibinfo{year}{2001}): \emph{\bibinfo{title}{Lookahead hierarchies of
  restarting automata}}.
\newblock {\slshape \bibinfo{journal}{Journal of Automata, Languages and
  Combinatorics}} \bibinfo{volume}{6}, pp. \bibinfo{pages}{493--506},
  \doi{10.25596/jalc-2001-493}.

\bibitemdeclare{incollection}{Nag2007}
\bibitem{Nag2007}
\bibinfo{author}{B.~\surnamestart Nagy\surnameend} (\bibinfo{year}{2008}):
  \emph{\bibinfo{title}{On $5'\to 3'$ sensing {W}atson-{C}rick automata}}.
\newblock In \bibinfo{editor}{M.\ \surnamestart Garzon\surnameend} \&
  \bibinfo{editor}{H.~\surnamestart Yan\surnameend}, editors: {\slshape
  \bibinfo{booktitle}{DNA Computing, 13th Intern. Meeting, Revised Selected
  Papers}}, {\slshape \bibinfo{series}{Lecture Notes in Computer Science}}
  \bibinfo{volume}{4848}, \bibinfo{publisher}{Springer},
  \bibinfo{address}{Heidelberg}, pp. \bibinfo{pages}{256--262},
  \doi{10.1007/978-3-540-77962-9_27}.

\bibitemdeclare{incollection}{NagKov14}
\bibitem{NagKov14}
\bibinfo{author}{B.\ \surnamestart Nagy\surnameend} \&
  \bibinfo{author}{L.~\surnamestart Kov{{\'a}}cs\surnameend}
  (\bibinfo{year}{2014}): \emph{\bibinfo{title}{Finite Automata with
  Translucent Letters Applied in Natural and Formal Language Theory}}.
\newblock In \bibinfo{editor}{N.T.\ \surnamestart Nguyen\surnameend},
  \bibinfo{editor}{R.\ \surnamestart Kowalczyk\surnameend},
  \bibinfo{editor}{A.\ \surnamestart Fred\surnameend} \&
  \bibinfo{editor}{F.~\surnamestart Joaquim\surnameend}, editors: {\slshape
  \bibinfo{booktitle}{Transactions on Computational Collective Intelligence
  {XVII}}}, {\slshape \bibinfo{series}{Lecture Notes in Computer Science}}
  \bibinfo{volume}{8790}, \bibinfo{publisher}{Springer},
  \bibinfo{address}{Heidelberg}, pp. \bibinfo{pages}{107--127},
  \doi{10.1007/978-3-662-44994-3_6}.

\bibitemdeclare{incollection}{otto176}
\bibitem{otto176}
\bibinfo{author}{B.\ \surnamestart Nagy\surnameend} \&
  \bibinfo{author}{F.~\surnamestart Otto\surnameend} (\bibinfo{year}{2010}):
  \emph{\bibinfo{title}{{CD}-systems of stateless deterministic {R(1)}-automata
  accept all rational trace languages}}.
\newblock In \bibinfo{editor}{A.H.\ \surnamestart Dediu\surnameend},
  \bibinfo{editor}{H.\ \surnamestart Fernau\surnameend} \&
  \bibinfo{editor}{C.~\surnamestart Martin-Vide\surnameend}, editors: {\slshape
  \bibinfo{booktitle}{{LATA 2010}, {P}roc.}}, {\slshape
  \bibinfo{series}{Lecture Notes in Computer Science}} \bibinfo{volume}{6031},
  \bibinfo{publisher}{Springer}, \bibinfo{address}{Berlin}, pp.
  \bibinfo{pages}{463--474}, \doi{10.1007/978-3-642-13089-2_39}.

\bibitemdeclare{incollection}{otto185}
\bibitem{otto185}
\bibinfo{author}{B.\ \surnamestart Nagy\surnameend} \&
  \bibinfo{author}{F.~\surnamestart Otto\surnameend} (\bibinfo{year}{2011}):
  \emph{\bibinfo{title}{Finite-state acceptors with translucent letters}}.
\newblock In \bibinfo{editor}{G.\ \surnamestart Bel-{E}nguix\surnameend},
  \bibinfo{editor}{V.\ \surnamestart Dahl\surnameend} \& \bibinfo{editor}{A.O.
  \surnamestart De~La~Puente\surnameend}, editors: {\slshape
  \bibinfo{booktitle}{{BILC 2011}: {AI} {M}ethods for {I}nterdisciplinary
  {R}esearch in {L}anguage and {B}iology, {P}roc.}},
  \bibinfo{publisher}{Sci{T}e{P}ress}, \bibinfo{address}{Portugal}, pp.
  \bibinfo{pages}{3--13}.

\bibitemdeclare{incollection}{NaOtLATA2011}
\bibitem{NaOtLATA2011}
\bibinfo{author}{B.\ \surnamestart Nagy\surnameend} \&
  \bibinfo{author}{F.~\surnamestart Otto\surnameend} (\bibinfo{year}{2011}):
  \emph{\bibinfo{title}{Globally deterministic {CD}-systems of stateless
  {R(1)}-automata}}.
\newblock In \bibinfo{editor}{A.H.\ \surnamestart Dediu\surnameend},
  \bibinfo{editor}{S.\ \surnamestart Inenaga\surnameend} \&
  \bibinfo{editor}{C.~\surnamestart Mart\'in-Vide\surnameend}, editors:
  {\slshape \bibinfo{booktitle}{Language and Automata Theory and Applications,
  {LATA 2011}, {P}roc.}}, {\slshape \bibinfo{series}{Lecture Notes in Computer
  Science}} \bibinfo{volume}{6638}, \bibinfo{publisher}{Springer},
  \bibinfo{address}{Berlin}, pp. \bibinfo{pages}{390--401},
  \doi{10.1007/978-3-642-21254-3_31}.

\bibitemdeclare{article}{otto195}
\bibitem{otto195}
\bibinfo{author}{B.\ \surnamestart Nagy\surnameend} \&
  \bibinfo{author}{F.~\surnamestart Otto\surnameend} (\bibinfo{year}{2012}):
  \emph{\bibinfo{title}{On {CD}-systems of stateless deterministic {R}-automata
  with window size one}}.
\newblock {\slshape \bibinfo{journal}{\JCSS}} \bibinfo{volume}{78}, pp.
  \bibinfo{pages}{780--806}, \doi{10.1016/j.jcss.2011.12.009}.

\bibitemdeclare{article}{otto206}
\bibitem{otto206}
\bibinfo{author}{B.\ \surnamestart Nagy\surnameend} \&
  \bibinfo{author}{F.~\surnamestart Otto\surnameend} (\bibinfo{year}{2013}):
  \emph{\bibinfo{title}{Globally deterministic {CD}-systems of stateless
  {R}-automata with window size~1}}.
\newblock {\slshape \bibinfo{journal}{\it International Journal of Computer
  Mathematics}} \bibinfo{volume}{90}, pp. \bibinfo{pages}{1254--1277},
  \doi{10.1080/00207160.2012.688820}.

\bibitemdeclare{incollection}{otto139}
\bibitem{otto139}
\bibinfo{author}{F.~\surnamestart Otto\surnameend} (\bibinfo{year}{2006}):
  \emph{\bibinfo{title}{Restarting automata}}.
\newblock In \bibinfo{editor}{Z.\ \surnamestart {{\'E}}sik\surnameend},
  \bibinfo{editor}{C.\ \surnamestart Mart{\'{i}}n-{V}ide\surnameend} \&
  \bibinfo{editor}{V.~\surnamestart Mitrana\surnameend}, editors: {\slshape
  \bibinfo{booktitle}{Recent {A}dvances in {F}ormal {L}anguages and
  {A}pplications}}, {\slshape \bibinfo{series}{Studies in {C}omputational
  {I}ntelligence}}~\bibinfo{volume}{25}, \bibinfo{publisher}{Springer},
  \bibinfo{address}{Heidelberg}, pp. \bibinfo{pages}{269--303}.

\end{thebibliography}

\end{document}